\documentclass[12pt]{article}
\def\cdate{{November 10, 2008}}
\usepackage{
amsfonts,amsmath,latexsym,amssymb,
txfonts,bm}
\usepackage[american]{babel}
\usepackage[dvips]{graphicx}

\def\II{{\mathbb I}}

\def\tr{\mathrm{ tr\,}}

\def\be{\begin{equation}}
\def\ee{\end{equation}}
\def\bea{\begin{eqnarray}}
\def\eea{\end{eqnarray}}
\def\bed{\begin{definition}{\ }}
\def\eed{\end{definition}}
\def\bd{\begin{description}}
\def\ed{\end{description}}
\def\bc{\begin{center}}
\def\ec{\end{center}}

\newtheorem{definition}{Definition}

\def\sideremark#1{\ifvmode\leavevmode\fi\vadjust{\vbox to0pt{\vss
\hbox to 0pt{\hskip\hsize\hskip1em
\vbox{\hsize2cm\tiny\raggedright\pretolerance10000
\noindent #1\hfill}\hss}\vbox to8pt{\vfil}\vss}}}

\makeatletter
\def\timenow{
\@tempcnta=\time \divide\@tempcnta by 60
\number\@tempcnta:\multiply \@tempcnta by 60 \@tempcntb=\time
\advance\@tempcntb by -\@tempcnta \ifnum\@tempcntb <10
0\number\@tempcntb\else\number\@tempcntb\fi}
\newcounter{outputpage}
\renewcommand{\@oddhead}
{\stepcounter{outputpage}\hfill\hfill\theoutputpage}
\renewcommand{\@evenhead}
{\stepcounter{outputpage}\hfill\hfill\theoutputpage}
\renewcommand{\@oddfoot}
{\vbox{
\vspace{3pt} \hfil {\scriptsize\textit{
\hfill\hfill
}}
\hfil }}
\renewcommand{\@evenfoot}
{\vbox{
\vspace{3pt} \hfil {\scriptsize\textit{
\hfill\hfill
}}
\hfil }}
\makeatother

\begin{document}

\begin{titlepage}
\thispagestyle{empty} \null

\hspace*{50truemm}{\hrulefill}\par\vskip-4truemm\par
\hspace*{50truemm}{\hrulefill}\par\vskip5mm\par
\hspace*{50truemm}{{\large\sc New Mexico Tech {\rm
(\cdate)}}}\vskip4mm\par
\hspace*{50truemm}{\hrulefill}\par\vskip-4truemm\par
\hspace*{50truemm}{\hrulefill}
\par
\bigskip
\bigskip
\par
\par
\vspace{1cm}
\centerline{\huge\bf A Model for the Pioneer Anomaly}
\bigskip
\bigskip
\centerline{\huge\bf}
\bigskip
\bigskip
\centerline{\Large\bf Ivan G. Avramidi and Guglielmo Fucci}
\bigskip
\centerline{\it New Mexico Institute of Mining and Technology}
\centerline{\it Socorro, NM 87801, USA}
\centerline{\it E-mail:
iavramid@nmt.edu, gfucci@nmt.edu}
\bigskip
\medskip
\vfill

{\narrower
\par In a previous work we showed that massive test particles exhibit a
non-geodesic acceleration in a modified theory of gravity obtained by a
non-commutative deformation of General Relativity (so-called Matrix
Gravity). We propose that this non-geodesic acceleration might be the
origin of the anomalous acceleration experienced by the Pioneer 10 and
Pioneer 11 spacecrafts.

\par}

\vfill

\end{titlepage}

\section{Introduction}
\setcounter{equation}{0}

The Pioneer anomaly has been studied by many authors (see
\cite{anderson98, anderson02,moffat04,reynaud06,jaekel06,tangen07}
and the references in these papers) and it has a pretty strong
experimental status \cite{laemmerzahl06}. It exhibits itself in an
anomalous acceleration of the Pioneer 10 and 11 spacecrafts in the
range of distances between $20 {\rm AU}$ and $50 {\rm AU}$ ($\sim
10^{14}{\rm cm}$) from the Sun. The acceleration is directed
toward the Sun and has a magnitude of \cite{anderson98, anderson02}
\be A^r_{\rm anom}\approx (8.74\pm1.33)\times  10^{-8} {\rm cm/s}^2 \,.
\ee
In the last
years there have been many attempts to explain the Pioneer anomaly
by modifying General Relativity (see, for example,
\cite{moffat04} and the references therein). However, there is
also some evidence \cite{tangen07} that it could not be explained
within standard General Relativity since it exhibits a {\it
non-geodesic motion}. That is, it cannot be explained by just
perturbing the Schwarzschild metric of the Solar system. It seems,  from
the analysis of the trajectories, that the spacecrafts do not  move
along the geodesics of any metric. Another puzzling fact is  that there
is no measurable anomaly in the motion of the planets  themselves, which
violates the equivalence principle. In other  words, the heavy objects
like the planets, with masses greater  than $\sim 10^{27}{\rm g}$, do
not feel any anomaly while the smaller objects, like the Pioneer
spacecrafts, with masses of  order $\sim 10^{5}{\rm g}$, do experience
it.

There are also some interesting numerical coincidences regarding the
Pioneer  anomaly (noticed in \cite{Makela07} as well). Recall that the
cosmological distance, which can be defined either by the Hubble
constant $H$ or by the cosmological constant $\Lambda$, is of  order
\be\label{1}
r_0\sim \frac{c}{H} \sim
\frac{1}{\sqrt{\Lambda}} \sim 10^{28} {\rm cm}
\ee
and the Compton
wavelength of the proton is of order
\be
r_1\sim \frac{\hbar}{m_p
c}\sim 10^{-13} {\rm cm}\,.
\ee
Now, we easily see, first of all,
that there is the following numerical relation
\be
\left(\frac{r_1}{r_{\rm anom}}\right) \sim \left(\frac{r_{\rm
anom}}{r_0}\right)^2\,,
\ee
where $r_{\rm anom}\sim 10^{14}{\rm
cm}$ is the distance at which the anomaly is observed. This means
that
\be
r_{\rm anom}\sim
\left(\frac{\hbar}{m_pc\Lambda}\right)^{1/3} \sim
\left(\frac{\hbar c}{m_p H^2}\right)^{1/3}\;.
\label{14xxx}
\ee
Secondly, the
characteristic distance determined by the value of the anomalous
acceleration, $A_{\rm anom}\sim 10^{-8} {\rm cm/sec}^2$, is of the
same order as the cosmological distance
\be
r_2\sim
\frac{c^2}{A_{\rm anom}}\sim 10^{28} {\rm cm}\,,
\ee
which simply
means that
\be
A_{\rm anom}\sim Hc\sim c^2\sqrt{\Lambda}\,.
\ee
It is very intriguing to speculate that the {\it Pioneer effect is
the result of some kind of interplay between the microscopic and
cosmological effects at the macroscopic scales}.

In this paper we apply the investigation of  motion of test particles in
an extended  theory of gravity, called Matrix Gravity, initiated in
\cite{avramidi08} to study the anomalous acceleration of Pioneer 10 and
Pioneer 11 spacecrafts. Matrix Gravity was proposed in a series of
recent papers
\cite{avramidi03,avramidi04a,avramidi04b}.
This is a modification of the standard
General Relativity in which the metric tensor $g^{\mu\nu}$
is replaced by a Hermitian $N\times N$
{\it matrix-valued symmetric two-tensor}
$
a^{\mu\nu}=g^{\mu\nu}\II+h^{\mu\nu}\,,
$
where $\II$ is the identity matrix,
$h^{\mu\nu}$ is a matrix-valued traceless symmetric tensor, i.e.
$\tr h^{\mu\nu}=0$.
In this theory the
usual interpretation of gravity as Riemannian geometry is no longer
appropriate. Instead, Matrix Gravity leads, quite naturally, to
a generalized geometry, that we call Matrix Geometry, which is
equivalent to a collection of Finsler geometries.
Instead of a usual Riemannian  geodesic flow, we get a system of
Finsler flows,
and, moreover, the mass of a test particle is replaced by
a  collection of mass parameters. In the commutative limit, only the
total mass is observed. For more details and discussions see
\cite{avramidi04a,avramidi04b,avramidi08}.

The dynamics of the tensor field $a^{\mu\nu}$ is described by a
non-commutative Einstein-Hilbert action, which can be constructed either
by an extension  of all standard concepts of differential geometry to
the  non-commutative setting \cite{avramidi03,avramidi04a} or from the
spectral invariant of a  partial differential operator of non-Laplace
type \cite{avramidi04b}.

The main goal of the present paper is to apply our previous study
\cite{avramidi08} of the motion of  test particles (in a simple model of
matrix gravity) to the Pioneer anomaly. We would like to stress that this study is just a first
attempt to analyze the phenomenological  effects of Matrix Gravity. We
do not claim that this simple model definitely solves the mystery of the
anomaly. Our aim is just to  propose another candidate for its origin.
Only future tests and  more detailed models can describe
the Pioneer anomaly in full  capacity. This work does not represent the
final answer, but  just a first attempt of studying this phenomenon
within the  framework of Matrix Gravity.

\section{Anomalous Acceleration in Matrix Gravity}
\setcounter{equation}{0}


In Matrix Gravity a massive particle is decsribed not by a single mass
parameter $m$ but rather by  $N$ different mass  parameters $m_i$, so
that  $m=\sum_{i=1}^N m_i$. In the
commutative limit we only observe the total mass
$m$.  The interesting question of the physical origin of the parameters
$m_i$  requires further study. For this reason, we do not assume that the $m_i$ are
positive. Following \cite{avramidi08} we consider  two different cases.
In the first case, that we  call the {\it nonuniform model}, we assume
that mass parameters are  different, and in the second case, that we
call the {\it uniform model},  we discuss what happens if they are equal
to each other, that is, $m_i=m/N$.

The equations of motion of a test
particle are derived and studied in \cite{avramidi08}. They have the
form
\be
\frac{d^2 x^\mu}{dt^2}
+\gamma^\mu{}_{\alpha\beta}(x,\dot x)\dot x^\alpha \dot x^\beta
=0\,,
\ee
where $\gamma^\mu{}_{\alpha\beta}(x,\dot x)$ are generalized
Christoffel coefficients that are homogeneous functions of $\dot x$
of order zero, in other words, they depend on the direction of $\dot x$,
but not on its magnitude. These functions
depend in a complicated way on the matrix-valued metric
$a^{\mu\nu}$, on the velocity, $\dot x^\mu$,  and, in general, on the
ratios $\mu_i=m_i/m$.

In the perturbation theory, when one writes
$a^{\mu\nu}=g^{\mu\nu}\II+h^{\mu\nu}$, the generalized Christoffel
coefficients are
\be
\gamma^\mu{}_{\alpha\beta}(x,\dot x)=
\Gamma^\mu{}_{\alpha\beta}(x)
+\theta^\mu{}_{\alpha\beta}(x,\dot x)\,,
\ee
where $\Gamma^\mu{}_{\alpha\beta}$ are usual
Christoffel coefficients of the metric of the metric $g_{\mu\nu}$
and $\theta^\mu{}_{\alpha\beta}(x,\dot x)$ is some tensor of first order
in the perturbation.
Now, the matrix-valued metric $a^{\mu\nu}=g^{\mu\nu}\II+h^{\mu\nu}$
satisfies the non-commutative Einstein equations derived in
\cite{avramidi04a,avramidi04b,fucci07}. As a result of non-commutative
corrections even the equations for the metric $g^{\mu\nu}$ are modified.
This means that both the metric $g^{\mu\nu}$ and the Christoffel symbols
$\Gamma^\mu{}_{\alpha\beta}$ are modified. More precisely, we let
\be
g^{\mu\nu}=g^{\mu\nu}_0+\beta^{\mu\nu},
\ee
where $g^{\mu\nu}_0$
is the non-perturbed Riemannian metric given by the solution of the
standard Einstein equations without any non-commutative corrections
and $\beta^{\mu\nu}$ is the non-commutative correction.
Then
\be
\Gamma^\mu{}_{\alpha\beta}=\Gamma_0^\mu{}_{\alpha\beta}
+\alpha^\mu{}_{\alpha\beta}\,,
\ee
where $\Gamma_0^\mu{}_{\alpha\beta}$ is the Christoffel symbols
of the metric $g_{0}^{\mu\nu}$ and $\alpha^\mu{}_{\alpha\beta}$
is the perturbation.

Thus, the equations of motion take the form
\be
\frac{d^2 x^\mu}{dt^2}
+\Gamma^\mu_0{}_{\alpha\beta}(x)\dot x^\alpha \dot x^\beta
=A^\mu_{\rm anom}(x,\dot x)\,,
\ee
where
\be
A^\mu_{\rm anom}(x,\dot x)=A^\mu_{\rm geod}(x,\dot x)
+A^\mu_{\rm non-geod}(x,\dot x)
\ee
is the anomalous acceleration and
and
\be
A^\mu_{\rm geod}(x,\dot x)=-\alpha^\mu{}_{\alpha\beta}(x)
\dot x^\alpha \dot x^\beta\,,
\ee
\be
A^\mu_{\rm non-geod}(x,\dot x)=-\theta^\mu{}_{\alpha\beta}(x,\dot x)
\dot x^\alpha \dot x^\beta\,.
\ee
are the geodesic and non-geodesic parts of the anomalous acceleration.
It is this anomalous acceleration that we are
going to study in this paper. We suggest that this
might explain the anomalous behavior of Pioneer 10 and 11 spacecrafts.

The anomalous geodesic acceleration can be easily computed by expanding
the Christoffel coefficients in the perturbation.
In the first order in $\beta$ we obtain
\be
A^\mu_{\rm geod}=
\frac{1}{2}\left(2\nabla_\alpha\beta^\mu{}_\beta
-\nabla^\mu\beta_{\alpha\beta}
\right)\dot x^\alpha\dot x^\beta\,,
\ee
where the covariant derivatives and all tensor operations are
performed with the non-perturbed metric $g^{\mu\nu}_0$.

The anomalous non-geodesic
acceleration was derived within perturbation theory in the
deformation parameter in \cite{avramidi08}. We study the two cases
mentioned above.

{\it Nonuniform Model.}
First, we study the generic case when the parameters
$\mu_i$ are different.
We define a function
\be
P(x,\xi)=\sum_{i=1}^N\mu_i\lambda_i(x,\xi)\,,
\ee
where $\xi_\mu$ is a covector and
$\lambda_i(x,\xi)$ are the eigenvalues of the matrix
$h^{\mu\nu}(x)\xi_\mu\xi_\nu$.
Note that since $\tr h^{\mu\nu}=0$ the matrix $h^{\mu\nu}\xi_\mu\xi_\nu$
is traceless, which implies that the sum of its eigenvalues is equal to
zero. Thus, in the uniform case, when all mass parameters $\mu_i$ are
the same, the function $P(x,\xi)$ vanishes. In this case the effects of
non-commutativity are of the second order.

The non-geodesic acceleration was computed in \cite{avramidi08}
and has the form
\be
A^\mu{}_{\rm non-geod}=
\frac{1}{2}
g^{\mu\nu}\left(
2\nabla_\alpha q_{\beta\nu}(x,\dot x)
-\nabla_\nu q_{\alpha\beta}(x,\dot x)
\right)\dot x^\alpha \dot x^\beta\,,
\ee
where
\be
q^{\mu\nu}(x,\xi)
=\frac{1}{2}\frac{\partial^2}{\partial \xi_\mu\partial\xi_\nu}
P(x,\xi)\,
\label{275xx}
\ee
and
the covariant derivatives are defined with the Riemannian metric.
Thus, the total anomalous acceleration is
\be
A^\mu{}_{\rm anom}=
\frac{1}{2}
\left[
2\nabla_\alpha (\beta^\mu{}_\beta+q^\mu{}_{\beta})
-\nabla^\mu (\beta_{\alpha\beta}+q_{\alpha\beta})
\right]\dot x^\alpha \dot x^\beta\,.
\ee

{\it Uniform Model.}
Now, we will simply assume that all mass parameters
are equal, that
is,
$
m_i=\frac{m}{N}\,.
$
The non-geodesic acceleration, computed in \cite{avramidi08},
has the form
\be
A^\mu{}_{\rm non-geod}=
-\frac{1}{8}g^{\mu\nu}\left(
2\nabla_\alpha S_{\beta\nu\rho\sigma}
-\nabla_\nu S_{\alpha\beta\rho\sigma}
\right)
\dot x^\rho\dot x^\sigma\dot x^\alpha\dot x^\beta\,,
\label{283xx}
\ee
where
\be
S^{\mu\nu\alpha\beta}=\frac{1}{N}\tr
(h^{\mu\nu}h^{\alpha\beta})\,.
\ee
Thus, the total anomalous acceleration is
\be
A^\mu{}_{\rm anom}=
\frac{1}{2}
\left(
2\nabla_\alpha \beta^\mu{}_\beta
-\nabla^\mu \beta_{\alpha\beta}
\right)\dot x^\alpha \dot x^\beta
-\frac{1}{8}g^{\mu\nu}\left(
2\nabla_\alpha S_{\beta\nu\rho\sigma}
-\nabla_\nu S_{\alpha\beta\rho\sigma}
\right)
\dot x^\rho\dot x^\sigma\dot x^\alpha\dot x^\beta\,.
\ee

In the spherically symmetric background, in the
non-relativistic limit, the radial anomalous acceleration
is given by:
in the uniform model,
\be
A^r{}_{\rm anom}=\frac{\partial}{\partial r}
\left(-\frac{1}{2}\beta^{00}(r)
+\frac{1}{8} S^{0000}(r)\right)
\,,
\ee
and, in the nonuniform model,
\be
A^r{}_{\rm anom}=\frac{\partial}{\partial r}
\left(
-\frac{1}{2}\beta^{00}(r)
-\frac{1}{2}q^{00}(r)\right)
\,.
\ee
Of course, this can also be interpreted
as a modification of Newton's Law \cite{avramidi08}.

Here, of course, the tensor components $\beta^{00}$, $S^{0000}$ and $q^{00}$
should be obtained as solution of the non-commutative
Einstein field equations (in the perturbation theory).
These equations are somewhat complicated. That is why, in this paper
we consider a {\it toy model} just to get a glimpse into the phenomenon.

We consider a simple model of $2\times 2$ real symmetric
commutative matrices. The static spherically symmetric
solution of the matrix
Einstein equations for this model
was obtained in \cite{avramidi08}. In the spherical coordinates
$x^0=t$, $x^1=r$, $x^2=\theta$, $x^3=\varphi$ it has the form
\begin{eqnarray}\label{29xxx}
a^{00}&=& A\;,\qquad
a^{11}=B\;,
\\[10pt]
a^{22}&=&\frac{1}{r^2}\;\mathbb{I}\;, \qquad
a^{33}=\frac{1}{r^{2}\sin^{2}\theta} \;\mathbb{I}\;,
\nonumber
\end{eqnarray}
with
\bea
B(r)&=&\left(1-\frac{1}{3}\Lambda r^2-\frac{r_g-\theta L}{r}\right)\II
+\frac{L}{r}\tau\,,
\nonumber\\[12pt]
A(r)&=&\varphi(r)\II+\psi(r)\tau\,,
\eea
where $\II=\left(%
\begin{array}{cc}
  1 & 0 \\
  0 & 1 \\
\end{array}%
\right) $, $\tau=\left(%
\begin{array}{cc}
  0 & 1 \\
  1 & 0 \\
\end{array}%
\right)\;$,
$\Lambda$ is the cosmological constant, $r_g=2GM$, $M$
is the mass of the Sun, $\theta$ and $L$ are some integration
parameters, (the parameter $\theta$ should not be confused with the
angle variable
$\theta$ above), and $\varphi(r)$ and $\psi(r)$ are some functions
(the function $\varphi(r)$ should not be confused with the angle
variable $\varphi$ above).
The functions $\varphi(r)$ and $\psi(r)$ can be parametrized
by
\be\label{2}
\varphi(r)=-f_1(r)+2\theta f_2(r),\qquad
\psi(r)=\theta f_1(r)+(1+\theta^2)f_2(r)\,.
\ee
By introducing the following function
\be
u(r)=1-\frac{1}{3}\Lambda r^2-\frac{r_g}{r}\,,
\ee
we obtain, for $f_{1}(r)$ and $f_{2}(r)$ in (\ref{2}), the expressions
\be
f_1(r)=\frac{u(r)}{\left[u(r)+(\theta+1)\frac{L}{r}\right]
\left[u(r)+(\theta-1)\frac{L}{r}\right]}\;,
\ee
\be
f_2(r)=\frac{\frac{L}{r}}{\left[u(r)+(\theta+1)\frac{L}{r}\right]
\left[u(r)+(\theta-1)\frac{L}{r}\right]}\;.
\ee


Notice that when the parameters $\theta$ and $L$ vanish the functions
$\varphi(r)$ and $\psi(r)$ give
nothing but the standard Schwarzschild solution with the cosmological constant \cite{avramidi08}.
We take the standard Schwarzschild solution
(without the cosmological constant)
as the non-perturbed
metric $g^{\mu\nu}_0$. The parameters $\theta$, $L$ and $\Lambda$ will be
considered as perturbation.
This means that the relevant components have the form
\be
\beta^{00}=\varphi(r)+\frac{1}{1-\frac{r_g}{r}}\,,
\ee
\be
h^{00}
=\psi(r)\tau\,,
\label{5158}
\ee
and, therefore,
\be
S^{0000}=\psi^2(r)\,.
\ee
In this simple $2\times 2$
model the tensor $q^{\mu\nu}$ is given by
$q^{\mu\nu}=\frac{\gamma}{2}\tr (h^{\mu\nu}\tau)$, \cite{avramidi08}, where
$\gamma=\mu_1-\mu_2$, and therefore,
\be
q^{00}=\gamma\psi(r)\,.
\ee

As we already mentioned above,
in the non-relativistic limit the only essential component
of the anomalous  acceleration is the radial one $A^r{}_{\rm anom}$.
By using the equations above we obtain: in the uniform model,
\bea
\label{333zz}
A^{r}_{\rm anom}
&=&
\frac{1}{2}\frac{\partial}{\partial r}\left[
-\frac{1}{1-\frac{r_g}{r}}
+f_1(r)-2\theta f_2(r)+\left(\frac{\theta}{2}f_1(r)
+(1+\theta^2)f_2(r)\right)^{2}
\right]\,,
\;\;\;\;\;\;\;
\eea
and in the non-uniform model,
\bea
\label{222zz}
A^{r}_{\rm anom}
&=&
-\frac{1}{2}\frac{\partial}{\partial r}\left[
\frac{1}{1-\frac{r_g}{r}}
+(\gamma\theta-1)f_1(r)+\gamma(1+\theta)^2f_2(r)
\right]
\,.
\eea

We would like to emphasize at this point  that the perturbation theory
is only  valid for small corrections. Obviously, when the corrections become large
one needs  to consider the exact equations of motion.

\section{Pioneer Anomaly}
\setcounter{equation}{0}

We have two free parameters in our model, ${\theta}$ and $L$
(and $\gamma$ in the non-uniform model).
We estimate these parameters to match the value of the observed
anomalous acceleration of the Pioneer spacecrafts.

First of all, we recall the
observed value of the cosmological constant
$\Lambda \approx 2.5 \cdot 10^{-56} {\rm cm}^{-2}$;
therefore,
$
r_0 \approx |\Lambda|^{-1/2}= 6.3 \cdot 10^{27} {\rm cm},
$
and the gravitational radius of the Sun
$
r_g\approx 1.5\cdot 10^5 {\rm cm}\,.
$
The relevant scale of the Pioneer anomaly is
$r_{\rm anom}\sim 10^{14}\div 10^{15} {\rm cm}$,
therefore, we can restrict our analysis to the range
$r_g<<r<<r_0$. The values of the dimensionless parameters are
$\frac{r_g}{r}\sim 10^{-8}$,
$\frac{r}{r_0}\sim 10^{-15}$, and
$\frac{r_g}{r_0} \sim 10^{-23}$.
We also remind that the value of the anomalous acceleration is
$
A^r_{\rm anom}\approx 8.7 \cdot 10^{-8} {\rm cm/s}^2 \,.
$
We should stress that
our  analysis only applies to the range of distances
relevant   for the study of the Pioneer anomaly. Therefore, strictly
speaking,  from a formal point of view, one cannot extrapolate our
equations beyond this interval.
Since the parameters  $\frac{r_g}{r}$, $\frac{r}{r_0}$ and
$\frac{r_g}{r_0}$  are negligibly small (compared to $1$)
they can be omitted.

By using the eqs. (\ref{333zz}) and (\ref{222zz}), and by defining
$\rho=(1+\theta^{2})L-\theta r_{g}$, we obtain \cite{avramidi08}
(in the usual units, $c$ being the speed of light)
for $r_g<<r<<r_0$: in the uniform model,
\bea
A^{r}{}_{\rm anom}
&=&-\frac{c^2}{4}
\left(\theta + \frac{\rho}{r}\right)
\left(\frac{\rho+2\theta r_g}{r^2}
-\frac{2}{3}\theta\Lambda r\right)\,,
\label{219xx}
\eea
and in the non-uniform model,
\bea
\label{222xzzb}
A^{r}_{\rm anom}
&=&\frac{c^2}{2}\gamma
\left(\frac{\rho+2\theta r_g}{r^2}
-\frac{2}{3}\theta\Lambda r
\right)\,.
\label{220xx}
\eea

{\it Uniform Model.}
First, we restrict to the case of vanishing cosmological constant.
Then the function
(\ref{219xx})
takes the form
\bea
A^{r}{}_{\rm anom}(r)
&=&-\frac{c^2}{4}
\left({\theta}+\frac{{\rho}}{r}\right)
\frac{{\rho+2\theta r_g}}{r^2}\,.
\eea
It has an extremum if the signs of
${\theta}$ and ${\rho}$ are different, which occurs at
$
r_*=-\frac{3}{2}\frac{{\rho}}{{\theta}}
$
and
is equal to
\be
A^r_{\rm anom}(r_*)=-\frac{c^2 {\theta}^3}{27}
\frac{(\rho+2\theta r_g)}{\rho^2}
\;.
\ee
Now, we assume that $r_*\sim r_{\rm anom}
\sim 10^{14}\,{\rm cm}$ and
$A^r_{\rm anom}(r_*)\sim
-10^{-8}{\rm cm}/{\rm sec}^2$
to
estimate the parameters
\be
{\rho}\sim 10^7 {\rm cm} \,,\qquad
{\theta}\sim -10^{-7}\,.
\ee

If we leave the cosmological constant there is another range of
parameters that should be investigated. Namely, when the term
$\frac{2{\theta}}{3 r_0^2}r$ becomes comparable with the term
$\frac{{\rho}}{r^2}$.
In this case the anomalous acceleration can be written, by
dropping negligible terms, as
\begin{equation}
A^{r}{}_{\rm anom}(r)=
-\frac{c^{2}}{4r_{0}}\left(\frac{{\theta}{\rho}
r_{0}}{r^{2}}
+\frac{2{\theta}^2}{3r_{0}} r\right)\;.
\end{equation}


We note that the term $\frac{c^{2}}{4r_{0}}$ gives the right magnitude of
the anomalous acceleration. If
we assume that the two terms in the parentheses
are comparable at the characteristic
length $r_{\rm anom}$ and are of order $1$, then we get
an estimate
\begin{equation}
{\rho}\sim
\frac{r_{\rm anom}^3}{r_{0}^{2}}{\theta}
\qquad\textrm{and}\qquad
{\theta} \sim
\left(\frac{r_{0}}{r_{\rm anom}}\right)^{\frac{1}{2}}\;,
\end{equation}
and, therefore,
\begin{equation}
{\rho}\sim 10^{-7}{\rm cm}\qquad\textrm{and}\qquad
{\theta}\sim
10^{7}\;.
\end{equation}

{\it Nonuniform Model.}
In the non-uniform model
we have an additional parameter $\gamma$. The function
has an extremum at
\be
r_*=\left(\frac{3{\rho} r_0^2}{{\theta}}\right)^{1/3}.
\ee
Now, we assume that $r_*\sim r_{\rm anom}\sim 10^{14}{\rm cm}$; then
\begin{equation}\label{10}
\frac{{\rho}}{{\theta}}
=
\frac{r_*^{3}}{3r_{0}^{2}}\sim
10^{-13} {\rm cm}\;.
\end{equation}
Further, by assuming $A^r{}_{\rm anom}(r_*)\sim
-10^{-8}{\rm cm}/{\rm sec}^2$ and using the eq.
we estimate the parameter $\gamma$
\begin{equation}
\gamma \sim 10^{13}\;.
\end{equation}

It is interesting to notice that, in this case, ${\rho}/\theta$ has the
same order of magnitude of the Compton wavelength of the proton.
Moreover, by using (\ref{10}), we confirm
the coincidence (\ref{14xxx})
mentioned in the
introduction.
This is very intriguing; it allows one to speculate
that the anomalous acceleration could be a result of an interplay
between the microscopic and macroscopic worlds, in other words,
the Pioneer anomaly could be a quantum effect.

\section{Conclusions}

In this paper we applied the kinematics of test particles
\cite{avramidi08} in Matrix Gravity \cite{avramidi04a,avramidi04b} to
the study of Pioneer anomaly.  Matrix Gravity is interpreted in terms of
Matrix Geometry, a generalized geometry which is equivalent to a
collection of Finsler geometries, rather than Riemannian geometry. This
new feature of our theory leads to an interesting and completely  new
phenomenon of splitting of Riemannian geodesics
to a collection of Finsler geodesics.
More precisely, instead of one  Riemannian metric we have different
Finsler metrics and different  mass parameters which describe the
tendency to follow a particular  trajectory determined by a particular
Finsler metric.  The interesting result is that test particles in our
theory  exhibit a non-geodesic motion which can be interpreted in
terms of  an anomalous acceleration.  This new feature led us to apply
these results for  studying the anomalous acceleration of the Pioneer
spacecrafts.

We considered two models: a uniform one, in which a particle
is described by a single mass parameter, and a non-uniform one, in which
a particle is described by multiple mass parameters.  The choice of one
model over the other should be dictated by physical reasons.  The
interesting question of whether the matter is described by only one mass
parameter or more than one mass parameters  requires further study. If
the Pioneer  anomaly is a new physical phenomenon we have to accept the
fact that  the equivalence principle does not hold. If this  is the
case, a model with different mass parameters  (violating the equivalence
principle) would be more  appropriate to describe the motion of test
particles in the Solar system.

The next step of our analysis of the phenomenological  consequences of
Matrix Gravity is to apply the kinematic model developed  in
\cite{avramidi08} to the study of galactic rotations. It would be  very
interesting to understand if the flat rotation curves of galaxies  can
be explained without the concept of dark matter.  We plan to investigate
this question in a future work.



\begin{thebibliography}{99}

\bibitem{anderson98}
Anderson J D, Laing P A, Lau E L, Liu A S, Nieto M M,  Turyshev S G 1998
Indication, from Pioneer 10/11, Galileo, and Ulysses data, of an
apparent anomalous, weak, long-range  acceleration, \emph{Phys. Rev.
Lett.} {\bf 81} 2858-2861

\bibitem{anderson02}
Anderson J D, Laing P A, Lau E L, Liu A S, Nieto M M,  Turyshev S G 2002
Study of the anomalous acceleration of Pioneer 10 and  11, \emph{Phys.
Rev.} D {\bf 65}, 082004

\bibitem{avramidi03} Avramidi I G 2003 A Noncommutative deformation of
General Relativity,
\emph{Phys. Lett. B} {\bf 576} 195--198

\bibitem{avramidi04a} Avramidi I G 2004 Matrix General Relativity: a
new look at
old problems, \emph{Class. and Quantum Grav.} {\bf  21} 103--120

\bibitem{avramidi04b} Avramidi I G 2004
Gauged gravity via spectral asymptotics of non-Laplace type operators,
\emph{J. High Energy Phys.} {\bf 07} 030

\bibitem{avramidi08} Avramidi I G and Fucci G 2008
Kinematics in Matrix Gravity, (to appear in
\emph{Gen. Rel. Grav.}), DOI: 10.1007/s10714-008-0713-6\\
arXiv:gr-qc/0802.3927

\bibitem{fucci07} Fucci G and Avramidi I G 2008
Noncommutative Einstein equations, \emph{Class. and Quantum Grav.}
{\bf 25} 025005

\bibitem{jaekel06}
Jaekel M-Th and Reynaud S 2006  Gravity tests and the Pioneer anomaly,
in \emph{Laser, clocks and drag-free: exploration of relativistic
gravity in space}, Eds. H. Dittus, C.Lammerzahl and S. Turyshev,
(Berlin: Springer) 193; arXiv:gr-qc/0511020

\bibitem{laemmerzahl06}
L\"ammerzahl C, Preuss O and Dittus H 2006 Is the physics within the
Solar system really understood? arXiv:gr-qc/060452

\bibitem{Makela07} M\"{a}kel\"{a} J. 2007
Pioneer anomaly: an interesting numerical coincidence,
arXiv:0710.5460v1

\bibitem{moffat04}
Moffat J W 2004  Modified gravitational theory and the Pioneer 10 and 11
spacecraft anomalous acceleration, arXiv:gr-qc/0405076

\bibitem{reynaud06}
Reynaud S and Jaekel 2006  M-T Long range gravity tests and the Pioneer
anomaly, arXiv:gr-qc/0610160

\bibitem{tangen07} Tangen K 2007 Could the Pioneer anomaly have a
gravitational origin?
\emph{Phys. Rev.} {\bf D76} 042005

\bibitem{weinberg72} Weinberg S 1972 \emph{Gravitation and Cosmology:
Principles and Applications of the General Theory of Relativity}
(John Wiley and Sons)


\end{thebibliography}
\end{document}